\documentstyle[12pt]{article}

\begin{document}

\baselineskip = 18pt plus 1pt minus 1pt

\title{Heavy Quark  S-- to P--Wave Transitions in a Consistent Quark Model}

\author{  A Wambach
\\
\makebox[5cm]{}
\\
{\em Theoretical Physics,}\\
{\em Department of Physics,} \\
{\em 1 Keble Road, OXFORD, OX1 3NP, England.} \\
\\
\\
OUTP -- 93 28P\\
\\
\\
}
\vspace{2cm}
\date{ November 1993}

\maketitle

\begin{abstract}
\noindent In Heavy Quark Effective Theory the
calculation of form--factors of higher excited states in a non--relativistic
framework suffers severe inconsistencies
due to  frame dependencies.
We show how careful inclusion of  the Wigner--rotation
of the   spin of the light quark and a
covariant description of the meson wave function  resolves these
inconsisitencies. We match this onto HQET and compute  the relevant
form factors  $\xi_{1/2}$ and $\xi_{3/2}$ to order $v^2$.
\end{abstract}
\pagestyle{empty}

\newpage
\pagenumbering{arabic}
\pagestyle{plain}

\section{INTRODUCTION}

Recently it has been shown \cite{close1,close2} that the quark model,
when employed consistently, is able to fit the spectroscopy as well as
to calculate dynamical quantities in accordance with experiment. This
has been done for S-- to S--wave transitions of heavy mesons in the
context of Heavy Quark Effective Theory (HQET) \cite{rev}.

In the present work we are going to extend the formalism of ref.\cite{close1}
to the production of  P--wave states in heavy hadron decay. In particular we
shall calculate the P--wave
form--factors of HQET, namely $\xi_{1/2}$ and $\xi_{3/2}$.

There are several reasons for considering transitions into excited states.
Experimental data on inclusive decays of heavy mesons, i.e. B--mesons,
is already available \cite{argus,cleo} and direct measurements of
$B \rightarrow D^{**}$ should be expected from the proposed
B--factories. To evaluate inclusive decay--properties one might employ
two theoretical possibilities, either referring to inclusive
calculations (as in ref.\cite{altarelli}), or summing over all
channels in an exclusive model. However it turns out that widely used
models such as the BSW
\cite{wirbel} or the KS model \cite{korner1}  do not include higher
excited  states: the only one currently available is the ISGW model
\cite{isgur2}. However the ISGW model has problems emperically; e.g.\
it predicts a contribution of higher excited D--mesons to the overall
decay of $B\rightarrow D, D^*, D^{**}$ of 13\% , while experiments
favour a value of 21\%. A better understanding of decays into higher
excited states is of need.

There has been a lot of effort invested to compute form--factors which
control S-- to S--wave decays in the HQET (\cite{close1},
\cite{neubert2} -- \cite{sadzikowski}). However calculations of
form--factors for decays into higher excited states do not appear very
frequently in the literature.

Ali et al. \cite{ali} used a non--relativistic quark model (NRQM) with a
harmonic oscillator wavefunction to calculate P--wave form--factors.
However their calculation inherits  several weaknessess. The
form--factors they computed are zero for $vv'=1$, whereas the
Bjorken--sum rule \cite{bjorken} requires a non--vanishing
form--factor in this limit. Furthermore  there are traditional
problems with non--relativistic models; firstly, non--relativistic
calculations often depend on the frame of the physical system,
 and as we will show in section two, this is true already to the zeroth
order in the velocity of the heavy quark. Secondly, it has been shown
in a recent work \cite{close1} that the successful calculation of the
S-- to S--wave form--factor only arises when the Wigner--rotation of
the spin of the recoiling light quark is included. The inclusion of the
Wigner--rotation as
being necessary for restoring the consistency of a non--relativistic
calculation has also been mentioned by others \cite{iddir}.

The present work deals with the application of  this matching
of quark models onto HQET  in the context of S-- to P--wave
transitions which will resolve the problem of frame--dependency in the NRQM.

 We proceed as follows: In section 2 we derive the formalism and
present  the calculation in the NRQM which is then to be compared with
the consistent calculation in  section 3. In section 4 we conclude and
discuss possible apllications.

\section{THE FORMALISM}

HQET makes use of the fact that the spin degrees of freedom of the
heavy quark are independent of the light quark. It is therefore
possible to derive wave--functions of heavy mesons where the heavy and
light spin--degrees of freedom are separated
\cite{falk2,balk,hussain}. In particular ref.\cite{balk} formulates
the covariant wavefunctions for higher excited heavy mesons:
\begin{equation}
\begin{array}{ccc}
0^{++} & = & -\sqrt{\frac{1}{3}}\frac{1}{2}(1+\not{\!
v})(\not{\!\tilde{k}}-\tilde{k}\cdot v) \\
\\
1_{1/2}^{+} & = &
\sqrt{\frac{1}{3}}\frac{1}{2}(1+\not{\!v})\gamma_{5} \not{\!\tilde{\epsilon}}
(\not{\!\tilde{k}}-\tilde{k}\cdot v) \\
 \\
\\
1_{3/2}^{+} & = &
\sqrt{\frac{1}{6}}\frac{1}{2}(1+\not{\!
v})\gamma_{5}(-\not{\!\tilde{\epsilon}}
(\not{\!\tilde{k}}-\tilde{k}\cdot v) + 3\tilde{k}\cdot\tilde{\epsilon})\\
\\
2^{++} & = & \frac{1}{2}(1+\not{\!v})\gamma^\mu
\tilde{\epsilon}_{\mu\nu}\tilde{k}^\nu\\
\\
\end{array}
\label{mesonmatrix1}
\end{equation}
with $\tilde{\epsilon}_{\mu}$ being the polarization vector of the spin one
state and $\tilde{\epsilon}_{\mu\nu}$ is the spin 2 polarization
tensor. $v_\mu$ is the velocity of the meson and $\tilde{k}_\mu$ is
the four--momentum of the light degrees of freedom. Note that
quantities with tilda describe quantities in a boosted system, whereas
quantities without tilda refer to the rest frame of the meson.

The authors of ref.\cite{hussain} derive equations similar to eq.(1), but with
$\tilde{k}^\mu_{\perp}$ instead of  $\tilde{k}^\mu$, where
$\tilde{k}^\mu _{\perp} = \tilde{k}^\mu - (\tilde{k} \cdot v)v^\mu$.
 In practice this is equivalent to our formalism because $v^\mu$
dotted into the expressions in eq.(1) yield zero, so
that effectively it does not make any difference if one chooses
$\tilde{k}^\mu $ or $\tilde{k}^\mu _{\perp}$.

We shall derive eq.(1) in an explicit quark model formalism, where we
follow the  derivation given by \cite{brodsky} as a limit to the
Bethe--Salpeter  wavefunction \cite{bethe} and which was successfully
applied  in the context of S-- to S--wave transitions \cite{close1}.
Similar results for the spin--structure of the wave--functions can be
found in \cite{kuhn,guberina}.

Consider one heavy quark with mass $m_Q$ and a light antiquark with
mass $m_q$. To construct a tensor we multiply the spinors: $M =
u\bar{v}$. By combining the relevant spin--components  the meson
wave--functions for the pseudoscalar and the vector state in  the
rest--frame are:
\begin{equation}
\begin{array}{ccc}
M_0(\vec{v}=0) & = & \frac{1}{2}(1+\gamma_0)\gamma_5 \\
\\
M_1(\vec{v}=0) & = & \frac{1}{2}(1+\gamma_0)\not{\!\epsilon}
\end{array}
\label{mesonmatrix2}
\end{equation}
or, in an arbitrary frame:
\begin{equation}
\begin{array}{ccc}
M_0(v) & = & \frac{1}{2}(1+\not{\!v})\gamma_5 \\
\\
M_1(v) & = & \frac{1}{2}(1+\not{\!v})\not{\!\tilde{\epsilon}}
\end{array}
\end{equation}

As shown in ref.\cite{close1} it is possible to preserve this
formalism and to add the contribution coming from the internal
momentum of the two quarks in the meson. This introduces a further
multiplicative factor $(m_q - \tilde{\not{\!k}})\phi(|\vec{k}|)$,
where $k$ is the momentum of the light antiquark which in the rest
frame of the meson we identify with the internal momentum. Note that a
similar factor for the
heavy quark of the form $(m_Q + \tilde{\not{\!k}})$ is supressed in
the infinite mass limit of the heavy quark. This will create
corrections in the case of a b--quark of about 10\%.  For higher excited  meson
states it is necessary to combine the Clebsch--Gordons coming from the
spin--coupling with those
from the L=1 orbital function. We do this in the rest frame of the
meson and then boost the meson to derive the form in an arbitrary frame.

Using eq.(2) and multiplying this expression with the appropriate
P--wave--function (which is proportional to $\vec{k}\cdot \vec{\eta}$
where $\vec{\eta}$ is the orbital angular momentum projection vector), the
meson takes the form:

\begin{equation}
M_p(v) \propto \left(\begin{array}{cc} 0&X\\
                                 0&0 \end{array}\right) \vec{k}\cdot
\vec{\eta}(m_q - \not{\!k})\\
\label{mesonmatrix3}
\end{equation}
where $X=-1$ for the pseudoscalar state $(0^{-+})$ and
$X=\vec{\sigma}\cdot\vec{\epsilon}$ for the vector state $(1^{--})$.
The last factor is the Wigner--rotation of the light--quark as
derived in ref.\cite{close1}.
Now identifying the appropriate Clebsch--Gordons yields the L=1 mesons
$(1^{+-}, 0^{++}, 1^{++}, 2^{++})$. In the rest frame they are:

\begin{equation}
\begin{array}{ccc}
1^{+-}  & = & - \frac{1+v_0\gamma_0}{2} \gamma_5
\vec{k}\cdot\vec{\epsilon}(m_q - \not{\!k})\\
\\
0^{++} &  = & \frac{1+v_0\gamma_0}{2} \sqrt{\frac{1}{3}} (k_0v_0-\not{\!k})
(m_q-\not{\!k}) \\
\\
1^{++} & = & \frac{i}{\sqrt{2}} \frac{1+v_0\gamma_0}{2}
\vec{\epsilon}\cdot(\vec{k}\times\vec{\sigma}) (m_q - \not{\!k})\\
\\
2^{++} & = & \frac{1+v_0\gamma_0}{2}\epsilon_{ab}\sigma^a k^b (m_q -
\not{\!k})\\
\\
\end{array}
\end{equation}

The dynamics of heavy-light systems are best described by $J.j$ coupling
rather than $L.S$ coupling of angular momenta and so the
physical axial-vector mesons  in
 the infinite mass limit are a mixture of the $1^{+-}$, $1^{++}$ states.
In the rest frame we used first $S_Q - S_q$ coupling and then
 $L-S$ coupling. Therefore the new states are:
\begin{equation}
\begin{array}{ccc}
1_{1/2}^{+}& =& \sqrt{\frac{1}{3}} 1^{+-} + \sqrt{\frac{2}{3}}
1^{++}\\
\\
1_{3/2}^{+}& =& \sqrt{\frac{2}{3}} 1^{+-} - \sqrt{\frac{1}{3}} 1^{++}\\
\end{array}
\end{equation}
Before boosting the spin--wavefunction and deriving the final form of
the meson wavefunction we should say something about the spatial part
of the wavefunction  which is employed here. We use harmonic
oscillator wave functions in the rest--frame of the meson with the
parameters   given  in \cite{isgur2}. In a moving system this
wavefunction (as a scalar) does not change, but we have to multiply it
by a non--covariant factor to keep the normalization correct (see
\cite{close1}  and also \cite{pene}). The only free parameter is
therefore the oscillator strength  scaled by the mass of the light quark.

With these remarks we are able to write down the wave functions for
the mesons in question:

\begin{equation}
\begin{array}{ccc}
\\
0^{++} & = & -\frac{1}{2} \sqrt{\frac{1}{3}} (1+\not{\!
v})(\not{\!\tilde{k}}-\tilde{k}\cdot v) (m_q-\not{\!\tilde{k}})
[2(m_q+\tilde{k}\cdot v)]^{-\frac{1}{2}}\frac{\tilde{k}\cdot
v}{\tilde{k_0}} \phi_p(k)\\
\\
1_{1/2}^{+} & = & \frac{1}{2} \sqrt{\frac{1}{3}}
(1+\not{\!v})\gamma_{5} \not{\!\tilde{\epsilon}}
(\not{\!\tilde{k}}-\tilde{k}\cdot
v) (m_q-\not{\!\tilde{k}}) [2(m_q+\tilde{k}\cdot v)]^{-\frac{1}{2}}
\frac{\tilde{k}\cdot v}{\tilde{k_0}} \phi_p(k)\\
\\
\\
1_{3/2}^{+} & = & \frac{1}{2} \sqrt{\frac{1}{6}}(1+\not{\!v})
\gamma_{5}(-\not{\!\tilde{\epsilon}}(\not{\!\tilde{k}}-\tilde{k}\cdot
v) + 3\tilde{k}\cdot \tilde{\epsilon})(m_q-\not{\!\tilde{k}})
[2(m_q+\tilde{k}\cdot v)]^{-\frac{1}{2}} \frac{\tilde{k}\cdot
v}{\tilde{k_0}} \phi_p(k)\\
\\
2^{++} & = & \frac{1}{2}(1+\not{\!v})\gamma^\mu
\tilde{\epsilon}_{\mu\nu}\tilde{k}^\nu(m_q-\not{\!\tilde{k}})
[2(m_q+\tilde{k}\cdot  v)]^{-\frac{1}{2}} \frac{\tilde{k}\cdot
v}{\tilde{k_0}} \phi_p(k)\\
\\
\end{array}
\label{meson2}
\end{equation}

To calculate transition matrix elements we employ the trace formalism
of ref.\cite{balk}. They observe that in general the
spin--wave function can be written in the form
\begin{equation}
\phi = \chi_\alpha \tilde{k}^\alpha A_p
\end{equation}
where in our case $A_p$ includes the contribution coming from the
Wigner--rotation and the wavefunction.
For S-- to P--wave transitions they then find for the transition element:
\begin{equation}
\begin{array}{ccc}
M_\mu & = & \langle M_p(v')|J_\mu^{V-A}|M_s(v) \rangle \\
\\
& = & {\rm Tr} \{ \bar{\chi_\alpha}' \gamma_\mu(1-\gamma_5) \chi
(F_j(y)v^\alpha+F'_{j}(y)\gamma^\alpha + F''(y)v^{'\alpha}) \}\\
\end{array}
\end{equation}
where $j=1/2$ and $3/2$ for the
multiplets $(0^+,1^+_{1/2})$ and $(1^+_{3/2},2^+)$ respectively and where
$y=v\cdot v'$. Note that $F''(y)v_\mu'$ contracts with the
spin--wavefunctions to zero, i.e. $\chi_\alpha v^{'\alpha}=0$.
Inserting the  spin--wave functions into this expression one derives
the structure for the HQET identical to the one  given in
ref.\cite{falk2} by identifying the form factors:
\begin{equation}
\begin{array}{ccc}
\xi_{1/2}(y) & = & (y+1) F_{1/2} (y) - 3 F'_{1/2}(y)\\
\\
\xi_{3/2}(y) & = & F_{3/2}(y)
\\
\end{array}
\end{equation}
To evaluate these form--factors it is therefore necessary to determine
$F$ and $F'$. Before we do this in the next chapter for the
consistent  quark model let us first discuss the solution in the NRQM.

The functions $F$, $F'$ and $F''$ measure the contribution of the
light quark to the meson--decay. In the case of a non--relativistic
wave--function this contribution cannot reveal a further spin--term,
so that the function in front of $\gamma^\alpha$ has to be zero: $F'=0$.
Furthermore the wave--functions in the NRQM do not distinguish
between $j=1/2$ and $j=3/2$. A first result is therefore:
\begin{equation}
\xi_{1/2}^{NRQM}(y) = (y+1) \xi_{3/2}^{NRQM}(y)
\end{equation}

Using harmonic oscillator wavefunctions (as in the spectroscopy studies
\cite{isgur2}) an explicit calculation in the
rest frame of the final state gives:
\begin{equation}
F(y=v_0) = \frac{m_q}{\sqrt{2}\beta} e^{-\frac{m_q^2(y-1)}{2\beta^2}}
\end{equation}
Here $\beta$ is the coupling strength which should be about 0.41 GeV
\cite{isgur2}.

One of the weaknesses of the NRQM is that results
differ in different frames. In order to investigate the inherent
uncertainty of the model we repeat the calculation in a different
frame, namely  the rest--frame of the meson {\it before} the decay.
The general form of the contribution of the light degrees of freedom
to the decay element in the NRQM is:
\begin{equation}
\int {\rm d}^3k  k_\mu
\phi^*_p(|\vec{k}+m_q\vec{v'}|)\phi_s(|\vec{k}|) = F(y) v_\mu + F''(y) v'_\mu
\end{equation}
where in the rest frame of the decaying meson $v=(1,\vec{0})$ and
$v'=(v'_0,\vec{v}')$. To determine $F$ one therefore has to calculate
the zero component of this expression. It is not clear in a NRQM what
to take as the zero
component of a four--momentum. We calculated $F$ for $k_0=m_q$ and
$k_0=\omega=\sqrt{m_q^2+\vec{k}^2}$ in the harmonic oscillator model
and used the
standard parameters \cite{isgur2}. This  then gives for the function
the following numerical results, where the evaluation from eq.(12) is
stated first:
\begin{equation}
\begin{array}{ccc}
\\
F(y)  & = & 0.57 - 0.18 (y-1) + O((y-1)^2)\\
\\
F_{`m_q'}(y) & = &  0.57 - 0.75  (y-1) +  O((y-1)^2)\\
\\
F_{`\omega_q'}(y) & = & 1.43 - 0.89 (y-1) +  O((y-1)^2)\\
\\
\end{array}
\end{equation}
The result depends heavily on the interpretation of the
energy--component of the four--momentum and in neither case  does this
coincide with the result in the rest--frame of the meson after the
decay. Consequently it is rather
difficult to trust any of these calculations.

A different problem occurs if one extracts the P--wave form--factors directly
from the ISGW model. This is done by identifying the appropriate form--factors
in the two formalisms (see Appendix C). In the context of S-- to S--wave
transitions it has already been argued \cite{amundson} that the ISGW model
shows an inconsistency in their prediction of the Isgur--Wise function to order
$(y-1)$, and the result in the present context is similar. The ISGW model
reveals that in a non--relativistic quark model $\xi_{1/2}(1) = 2
\xi_{3/2}(1)$, however the expansion in $(y-1)$ is not unique. In particular
the ISGW model predicts the following different forms for $\xi_{3/2}$ (see
Appendix C for the origin of these expressions):

\begin{equation}
\begin{array}{ccrc}
\xi_{3/2}(y) & \rightarrow &  & \frac{m_d}{\sqrt{2}\beta}e^{-\frac{m_d^2(y-1)}
{2\beta^2\kappa^2}} \\
\\
     & \rightarrow & \frac{1}{\kappa^2} &  \frac{m_d}{\sqrt{2}\beta}
e^{-\frac{m_d^2(y-1)}{2\beta^2\kappa^2}}\\
\\
 & \rightarrow & \frac{2}{1+y} & \frac{m_d}{\sqrt{2}\beta}
e^{-\frac{m_d^2(y-1)}{2\beta^2\kappa^2}}\\
\\
& \rightarrow & \frac{1}{1-(y-1)\frac{m_b}{3m_c+m_b}} &
\frac{m_d}{\sqrt{2}\beta} e^{-\frac{m_d^2(y-1)}{2\beta^2\kappa^2}}\\
\\
\end{array}
\end{equation}

The factor $\kappa$ which appears in these expressions was introduced
in ref.\cite{isgur2} to compensate for relativistic effects. Its value
is $\kappa = 0.7$.

We are  now going to show how these inconsistencies, namely the
frame--dependency and the uncertainty in the $(y-1)$ expansion
disappear when the quark model is employed consistently.

\section{THE FORM FACTORS IN THE CONSISTENT  MODEL}

In this section we shall  illustrate how the formalism established in
section 2 enables us to calculate S-- to P--wave transition form--factors
to  the first order in $(y-1)$.
The kinematics of this decay has been discussed in
\cite{close1,close3}. We  state their results for the particular case where
the final state  meson is at rest. Before   the decay it was moving
with the velocity $v_\mu$ in the direction of the z--axis,
$v_\mu=(v_0,0,0,v)$. The four--momenta of the light antiquark before
(unprimed) and after (primed) the decay satisfy the following conditions:

\begin{equation}
\begin{array}{ccl}
\\
\omega & = & \omega' v_0 - k'_z v\\
\\
k_{x(y)}  & = & k'_{x(y)}\\
\\
k_z & = & -\omega' v + k'_z v_o \\
\\
\end{array}
\label{kinematic}
\end{equation}

Following the preparation  given above, the S-wave to P-wave
transition form--factors in the unified formulation  have the following form:
\begin{equation}
\begin{array}{c}
\int {\rm d}^3 \tilde{k}{\rm d}^3 k' k'_\mu
(m_q-\not{\!k}')[(m_q+\omega' v_o-vk'_z)(m_q+\omega')]^{-\frac{1}{2}}
(\frac{\omega' v_0 - vk'_z }{\omega'})^{\frac{1}{2}}\times \\
\\
\frac{\sqrt{2}}{\beta}
(\beta^2 \pi)^{-\frac{3}{2}} e^{-\frac{k'^2}{2\beta^2}}
e^{-\frac{k^2}{2\beta^2}}\delta^3(\tilde{\vec{k}}-\vec{k}')
\\
\end{array}
\end{equation}
where we used explicit L=0 and L=1 quark model wave functions as in
\cite{isgur2}, but without the relativistic compensation factor. The
origin of  these terms is as follows:
 $(m_q - \not{\!k}')$ is the spin part coming from the
light quark, i.e the Wigner--rotation of the light quark. This
was contracted with an additional factor  $(m_q - \tilde{\not{\!k}})$
where we used that $\tilde{k}=k'$ because of four--momentum
conservation for the light quark. The factor $[(m_q+\omega'
v_o-vk'_z)(m_q+\omega')]^{-\frac{1}{2}}$ is the normalization factor
of the additional spin contribution (see eg.\ ref\cite{close1}) and
$(\frac{\omega' v_0 - vk'_z }{\omega'})^{\frac{1}{2}}$ is the already
mentioned overall normalization factor.\footnote{The authors of
ref.\cite{isgur2} retained different oscillator--strengths for the
wave--functions to account for different quark masses. In the
heavy--mass limit we use only one paramter $\beta$ and finally vary
this one in order to get a measure for the intrinsic uncertainty of
the model}

{}From now on we rescale the expression by a common factor $m_q$. We
retain $\omega'$,$k'$ and  $\beta$ as
parameters for the four--momentum and coupling strength to avoid
further indices, but note that these are now  dimensionless parameters.
 Expanding (17)  in $v$ and keeping only terms proportional
to $v$, $v^2$ and $v^3$ one derives an expression for the integrand of
the form:
\begin{equation}
k_\mu'(1-\omega'\gamma_0 + \vec{k}' \cdot \vec{\gamma})
[T_0+vk'_z(T_1+v^2T_4) + v^2(T_2+k_z^{'2}T_3)]  \\
\end{equation}
where the $T_i$ depend on $\omega'$ but, apart from $T_4$, not on
$k'_z$ (see  appendix A for further details).
This term is now to be matched to $Fv_\mu + F'\gamma_\mu + F''v'_\mu$.
In the frame we are working with, namely
$v=(v_0,0,0,v)$ and $v'=(1,\vec{0})$ it is rather simple to identify $F'$:
\begin{equation}
F' = \int {\rm d}^3 k' k^{'2}_x [T_0 + v^2(T_2+k_z^{'2}T_3)] \\
\end{equation}
The discussion is a little bit more involved if one is to determine
$F$. The fourth component of the expression has the form (leaving the
integral aside):
\begin{equation}
Fv + F'\gamma_z \rightarrow k^{'2}_z\gamma_z[T_0 +
v^2(T_2+k^{'2}_zT_3)] + vk'^2_z(1-\omega'\gamma_0)(T_1+v^2T_4)\\
\end{equation}
To evaluate this expression one has to use that terms proportional to
$\not{\!v}$ or$\not{\!v}'$ can be contracted with the spin--part
coming from the heavy meson, i.e. they contract to $-1$.
\begin{equation}
\begin{array}{ccc}
Fv & \rightarrow & (k^{'2}_z-k^{'2}_x) \gamma_z[T_0 +
v^2(T_2+k^{'2}_zT_3)] +  vk'^2_z(T_1+v^2T_4) -
vk'^2_z\omega'\gamma_0(T_1+v^2T_4)\\
\\
   & \rightarrow &
-v(k^{'2}_z-k^{'2}_x)(k_z^{'2}T_3)(\not{\!v}-v_0\not{\!v}' )
+ vk'^2_z(T_1+v^2T_4)(1-\omega' \not{\!v}')\\
\\
\end{array}
\end{equation}
and therefore:
\begin{equation}
F = \int {\rm d}^3 k' [ k^{'2}_z(T_1+v^2T_4)(1+\omega') - (v_0-1)
(k^{'2}_z-k^{'2}_x)(k^{'2}_zT_3)] \\
\end{equation}
Numerical evaluation of this expression gives:
\begin{equation}
\begin{array}{ccc}
F(y) & = & 1.14 - 2.20(y-1) + O((y-1)^2)\\
\\
F'(y) & = & 0.29 - 0.46(y-1) +  O((y-1)^2)\\
\\
\end{array}
\end{equation}
and  for the form--factors:
\begin{equation}
\begin{array}{ccc}
\xi_{1/2}(y) & = & 1.43 - 1.86 (y-1) +  O((y-1)^2)\\
\\
\xi_{3/2}(y) & = & 1.14 - 2.20 (y-1) +  O((y-1)^2)\\
\end{array}
\end{equation}

The calculation in the rest--frame of the meson {\it before} the decay
is similar though somewhat more laborious. As in the
non--relativistic case one has to determine first $F''$ and then
$F$ via the zero'th component of the expression in eq.(17). In this
case there is no ambiguity about what to take as the zero component of
the four--momentum: it has to be $\omega$. It is not difficult to see that
the numerical value for $F''$ in this frame will coincide with the one
for $F$ in the rest--frame of the meson after the decay
and that $F'$ will  not change in the two different frames. If we note
the components in the new frame with an subindex $n$ we get for $F_n$
(see appendix B):
\begin{equation}
F_n \rightarrow \omega(1+\omega)[T_0+v^{'2}(T_2+k_z^2T_3)]
-\frac{1}{2}v^{'2} \omega k_z^2T_1 + F'_n - (1+\frac{1}{2}v^{'2})F_n''\\
\end{equation}
Explicit calculation verifies that $F_n \equiv F$. This also provides
a consistency check on the normalization of  the wave--functions.

In Figure 1 we show the zero and first order terms for $F$ and
$F'$ in an expansion in $(y-1)$, which we denote by $F'$, $F$, $f'$ and  $f$,
for different coupling strengths.
{}.


\input epsf
\newpage
\vspace*{-2.5cm}
\epsfbox{form.ps}
\vspace*{-13cm}
 {\it{ {\rm [1]} The figure shows the dependence of the functions $F(y)$ and
$F'(y)$ on a variation  of the oscillator--strength divided by the light quark
mass. In this plot $F$ and $F'$ denote the zero component of $F(y)$ and $F'(y)$
in an $(y-1)$
 expansion. $f$ and $f'$ are the first order components of the same
 functions. Note the flatness of the curves over a 50\%
 variation of $\beta/m_q$.}}

\vspace{1cm}
Two features are worth mentioning:
\begin{tabbing}
(i) \= This calculation is  frame independent, in contrast to this
well known problem \\
\> in non--relativistic models\\
(ii) \> The result is insensitive to varying  the ratio
$\frac{\beta}{m_q}$ over a wide range  \\
\> of reasonable values, i.e a variation of $\beta/m_q$ over 12\% varies $F'$
($F$, $f'$, $f$) for \\
\> about 7\% (3\%, 1\%, 7\%) (see figure [1]).\\
\end{tabbing}

The final result with the error given by a 12\% variation over the
coupling strength is:
\begin{equation}
\begin{array}{ccc}
\xi_{1/2}(y) & = & (1.43 \pm 0.13) - (1.86 \pm 0.28) (y-1) + O((y-1)^2)\\
\\
\xi_{3/2}(y) & = & (1.14 \pm 0.04) - (2.20 \pm 0.16) (y-1) + O((y-1)^2)\\
\\
\end{array}
\end{equation}

For completeness we present the results of two further calculations of
the heavy quark S-- to P--wave form--factors. Methods which are
employed are   QCD sum--rules \cite{blok} and  the Bethe--Salpeter
formalism \cite{dai}. Both calculations have been performed for the
zero--recoil limit (y=1) and while the former suffer under large
errors the latter does not quote possible uncertainties in their
approach.  The
QCD--sum--rule calculation yielded $\xi_{1/2}(y=1) = 1.2 \pm 0.7$, $\xi_{3/2}$
was not determined. In the Bethe--Salpeter
approach the authors computed $\xi_{1/2}(y=1) = 0.73$  and $\xi_{3/2}
(y=1) = 0.76$.

\section{CONCLUSIONS}

The self--consistent calculation of S--to S--wave transitions in
ref.\cite{close1} has been extended to decays into higher--excited states. S--
to P--wave form factors are computed in this model which includes the
Wigner--rotation of the light quark. The results for the form--factors are no
longer frame dependent and
furthermore they stay relativly stable over a wide range of different
parameters. Both aspects encourage us  that this model calculation may
give  a reliable estimate of these functions.

One of the important results is the value of the form--factors at zero recoil.
As was seen in \cite{close1} in the context of S-- to S--wave
transitions , {\it ``the Wigner spin rotation
effectively decreases the probability for the final state pseudoscalar
to  overlap with the initial and hence the
form factor falls faster ...
 than when this effect is ignored.''}   Unitarity now tells us that in
order to compensate for the decreased probability to the S--states, the
heavy meson decay  branching ratio
to higher excited states should be larger when including
Wigner--rotation effects. This explicit calculation confirms this, which can be
seen by comparing the first line of eq.(14) (essentially eq.(12)) and our
results (eqs.(23), (24)).

Unitarity is formally summarized by the Bjorken sum rule \cite{bjorken},
which for the case of present interest is:
\begin{equation}
\rho^2 = \frac{1}{4} + \frac{1}{12} |\xi_{1/2}(y=1)|^2 +
\frac{2}{3}|\xi_{3/2}(y=1)|^2 + ...
\end{equation}
where the dots stand for contributions from higher excitations and the
inelastic continuum.  $\rho$ is the so--called charge
radius of the Isgur--Wise function. With the value of $\rho$
determined in this model in \cite{close1} the sum--rule yields:
\begin{equation}
1.42 = 0.25 + 0.17 + 0.87 +  ...
\end{equation}

The calculation presented here may have significant consequences for the decay
$B\rightarrow D^{**}$, where the rate is modified in the direction that data
appear to require. It is now worth studying how this comes about as a function
of momentum transfer/lepton--momentum. This is under investigation.

\vskip 0.3in

The author is indebted to F.E. Close and F. Hussain for discussions.
Support from the German Academic Exchange Service and the German Scholarship
Foundation is gratefully acknowlegded.

\vspace{5cm}
\appendix{APPENDIX A}
\\
To determine $T_0$ to $T_4$ we have to expand expression (17) to order
$v^3$ by using the kinematical contraints from (15). This gives:
\begin{displaymath}
\begin{array}{ccl}
T_0 & = & 1 \\
\\
T_1 & = & \frac{1}{2(1+\omega')} + \frac{\omega'}{\beta^2} -
\frac{1}{\omega'} \\
\\
T_2 & = &
-\frac{\omega'}{4(1+\omega')}-\frac{\omega^{'2}}{2\beta^2} + \frac{1}{4}\\
\\
T_3 & = &  \frac{3}{8(1+\omega')^2} - \frac{1}{2\beta^2} +
\frac{\omega^{'2}}{2\beta^4}
+\frac{\omega'}{2\beta^2(1+\omega')}-\frac{1}{8\omega^{'2}} -
\frac{1}{4\omega'(1+\omega')} - \frac{1}{2\beta^2}\\
\\
T_4 & = & [-\frac{3\omega'}{8(1+\omega')^2} + \frac{1}{8\omega'}
-\frac{\omega^{'3}}{2\beta^4} + \frac{\omega'}{\beta^2} +
\frac{1}{4(1+\omega')} -
\frac{\omega^{'2}}{2\beta^2(1+\omega')}] +k_z^{'3}[ \frac{5}{16(1+\omega')^3} -
\frac{1}{16\omega^{'3}} - \frac{3\omega'}{4\beta^4}    \\
\\
& &  + \frac{\omega^{'3}}{6\beta^6} -
\frac{1}{16\omega^{'2}(1+\omega')}  - \frac{1}{2\beta^2(1+\omega')} +
\frac{\omega^{'2}}{4\beta^4(1+\omega')} - \frac{3}{16\omega'(1+\omega')^2} +
\frac{1}{8\beta^2\omega'} - \frac{3\omega'}{8\beta^2(1+\omega')^2}]\\
\end{array}
\end{displaymath}
\\
All terms have to be understood as being multiplied with a common
factor $\frac{1}{1+\omega'}e^{({-\vec{k}^{'2}/\beta^2})}$.
\\
\newpage
\appendix{APPENDIX B}
\\
The general expression for the light quark contribution in the frame
where the meson is at rest before the decay is:
\begin{displaymath}
k_\mu ( 1-\omega\gamma_0 +\vec{k}\cdot\vec{\gamma})[ T_0 +
k_zv'(T_1+v^{'2}T_4) + v^{'2}T_2 + v^{'2}k_z^2 T_3]
\end{displaymath}
where $T_i$ $(i=1,...,4)$ is given in appendix A, but with the primed
quantities now being unprimed and vice versa.
This has to be indentified with $F_n v_\mu + F_n' \gamma_\mu + F_n'' v'_\mu$
for $v=(1,\vec{0})$ and $v'=(v'_0,0,0,v')$.
Considering the zero'th component of this expression we get:
\begin{displaymath}
\begin{array}{ccl}
F_n +F_n'\gamma_0+F_n''v_0' & \rightarrow & \omega(1-\omega \gamma_0)[T_0 +
v^{'2}(T_2+k_z^2T_3)]
+ \omega k_z^2\gamma_z v'(T_1+v^{'2}T_4)\\
\\
& \rightarrow &  \omega[T_0 + v^{'2}(T_2+k_z^2T_3)] - F_n''v'_0  - \omega
k_z^2  (T_1+v^{'2}T_4)  \not{\!v'}  + \\
\\
& & \{ -F_n'- \omega^2[T_0 + v^{'2}(T_2+k_z^2T_3)] + \omega
k_z^2 (T_1+v^{'2}T_4)v'_0 \}  \not{\!v}\\
\\
& & + F_n''v_o + F_n'\gamma_0\\
\end{array}
\end{displaymath}
Using the contraction properties of $\not{\!v}$ and $\not{\!v}'$ one derives
eq.(25).
\\
\\
\appendix{APPENDIX C}
\\
In HQET the decays of a heavy meson with quantum numbers $0^-$ into heavy
mesons with quantum numbers $0^+$, $1^+$, $1^+$ and $2^+$ are described in the
following way \cite{falk2,balk,hussain}: \\
\begin{displaymath}
\begin{array}{l l l l}
D_2^*  : & \frac{ \langle D_2^*(v',\epsilon)| \gamma_\mu |B(v) \rangle }
{\sqrt{M_B M_{D_2^*}}} & =  &  \xi_{3/2}(y) i \epsilon_{\mu \alpha \beta \gamma
}
\epsilon^{* \alpha \eta} v_\eta v^{'\beta} v^\gamma \\
\\
        & \frac{ \langle D_2^*(v',\epsilon)|\gamma_\mu \gamma_5| B(v) \rangle}{
 \sqrt{M_B M_{D_2^*}}} & = &  \xi_{3/2}(y)[(y+1)\epsilon^*_{\mu \alpha}
v^{\alpha} - \epsilon^*_{\alpha \beta} v^\alpha v^\beta v_\mu']\\
\\
\\
D_1^{3/2}  : & \frac{\langle D_1^{3/2} (v',\epsilon)| \gamma_\mu| B(v)\rangle}{
\sqrt{M_B M_{D_1^{3/2}}}} & = &  \frac{-1}{\sqrt{6}}\xi_{3/2}(y)[( y^2 -1)
\epsilon^*_\mu + 3(\epsilon^* \cdot v) v_\mu -(y-2)(\epsilon^* \cdot v)
v'_\mu]\\
\\
       & \frac{\langle D_1^{3/2} (v',\epsilon)| \gamma_\mu \gamma_5|
B(v)\rangle}{   \sqrt{M_B M_{D_1^{3/2}}}} & = &
\frac{-i}{\sqrt{6}}\xi_{3/2}(y)( y+ 1) \epsilon_{\mu \alpha \beta \gamma}
\epsilon^{*\alpha} v^{'\beta} v^\gamma\\
\\
\end{array}
\end{displaymath}
\begin{displaymath}
\begin{array}{l l l l}
D_1^{1/2} : & \frac{\langle D_1^{1/2} (v',\epsilon) |\gamma_\mu|B(v)\rangle}{
\sqrt{M_B M_{D_1^{1/2}}}} & = &  \frac{-1}{\sqrt{3}}\xi_{1/2}(y)[(1- y)
\epsilon^*_\mu + (\epsilon^* \cdot v) v'_\mu]\\
\\
     & \frac{\langle D_1^{1/2} (v',\epsilon) |\gamma_\mu \gamma_5|
B(v)\rangle}{   \sqrt{M_B M_{D_1^{1/2}}}} & = &  \frac{i}{\sqrt{3}}\xi_{1/2}(y)
\epsilon_{\mu \alpha \beta \gamma} \epsilon^{*\alpha} v^{'\beta} v^\gamma\\
\\
\\
D_0^*  : & \frac{\langle D_0^* (v',\epsilon) |\gamma_\mu \gamma_5| B(v)\rangle}
{\sqrt{M_B M_{D_0^*}}} & = & \frac{1}{\sqrt{3}}\xi_{1/2}(y) (v'-v)_\mu
\\
\\
\end{array}
\end{displaymath}

In the ISGW model, where the functions appearing are functions of the variable
$q^2 = (P_B - P_D^{**})^2$, one has the following general expressions for the
decays into L=1 mesons:\\
\begin{displaymath}
\begin{array}{l l l l}
^3P_2: & \langle D_2^*(p_D,\epsilon)| \gamma_\mu |B(p_B)\rangle & =  &   i
h(q^2) \epsilon_{\mu \alpha \beta \gamma }
\epsilon^{* \alpha \eta} p_{B \eta} (p_B + p_D)^{\beta} (p_B-p_D)^\gamma \\
\\
 & \langle D_2^*(p_D,\epsilon)| \gamma_\mu \gamma_5| B(p_B)\rangle  & = &
k(q^2) \epsilon^*_{\mu \alpha} p_B^{\alpha} + \epsilon^*_{\alpha \beta}
p_B^\alpha p_B^\beta * \\
\\
& & & [b_+(q^2)(p_B+p_D)_\mu +b_-(q^2)(p_B-p_D)_\mu ]\\
\\
\\
^3 P_1  : & \langle D_1^3 (p_D,\epsilon) |\gamma_\mu| B(p_B)\rangle & = &
l(q^2)  \epsilon^*_\mu + (\epsilon^* \cdot p_B) [c_+(q^2)(p_B+p_D)_\mu
+c_-(q^2)(p_B-p_D)_\mu] \\
\\
& \langle D_1^3 (p_D,\epsilon)| \gamma_\mu \gamma_5| B(p_B)\rangle & = &  i
q(q^2)  \epsilon_{\mu \alpha \beta \gamma} \epsilon^{*\alpha}(p_B +
p_D)^{\beta} (p_B-p_D)^\gamma\\
\\
\\
^1 P_1 : & \langle D_1^1 (p_D,\epsilon) |\gamma_\mu| B(p_B)\rangle & = &
r(q^2)  \epsilon^*_\mu + (\epsilon^* \cdot p_B) [s_+(q^2)(p_B+p_D)_\mu
+s_-(q^2)(p_B-p_D)_\mu] \\
\\
& \langle D_1^1 (p_D,\epsilon)| \gamma_\mu \gamma_5| B(p_B)\rangle & = &  i
v(q^2)  \epsilon_{\mu \alpha \beta \gamma} \epsilon^{*\alpha}(p_B +
p_D)^{\beta} (p_B-p_D)^\gamma\\
\\
\\
^3 P_0 : &  \langle D_0^3 (p_D,\epsilon) |\gamma_\mu \gamma_5| B(p_B)\rangle &
= &  u_+(q^2)(p_B+p_D)_\mu +u_-(q^2)(p_B-p_D)_\mu\\
\\
\end{array}
\end{displaymath}

Identifying the $D_1^{3/2}$ and the $D_1^{1/2}$ by means of eq.(6) one derives
the following constraints for the fourteen form--factors, or equivalently,  the
following predictions for the two HQET--form--factors (where we use
$M_{D^{**}}$ as the mass of the P--wave D--mesons):\\
\begin{displaymath}
\begin{array}{lll}
\xi_{3/2}(y) & = & 2M_B \sqrt{M_B M_{D^{**}}} h(q^2)\\
\\
             & = & \sqrt{\frac{M_B}{M_{D^{**}}}}\frac{1}{(1+y)}k(q^2)\\
\\
             & = & -M_B \sqrt{M_B M_{D^{**}}} (b_+(q^2) - b_-(q^2))\\
\\
             & = & \frac{-\sqrt{2}}{(y^2-1)\sqrt{M_B M_{D^{**}}}}
(\sqrt{2}r(q^2) - l(q^2))\\
\\
             & = & \frac{-M_B}{3} \sqrt{\frac{2 M_B}{ M_{D^{**}}}} (
\sqrt{2}(s_+(q^2) + s_-(q^2)) - c_+(q^2) - c_-(q^2))\\
\\
             & = & \frac{\sqrt{2 M_B M_{D^{**}}}}{(y-2)} ( \sqrt{2}(s_+(q^2) -
s_-(q^2)) - c_+(q^2) + c_-(q^2))\\
\\
             & = & \frac{2\sqrt{2 M_B M_{D^{**}}}}{(y+1)} (\sqrt{2} v(q^2) -
q(q^2))\\
\\
\\
\xi_{1/2}    & = & \frac{1}{(y-1)\sqrt{M_B M_{D^{**}}}} (r(q^2) + \sqrt{2}
l(q^2))\\
\\
             & = & - \sqrt{ M_B M_{D^{**}}} ( s_+(q^2) - s_-(q^2) +\sqrt{2}(
c_+(q^2) - c_-(q^2)))\\
\\
             & = & 2\sqrt{ M_B M_{D^{**}}} ( v(q^2) +\sqrt{2} q(q^2))\\
\\
             & = & -\frac{1}{2} \sqrt{\frac{3}{M_B M_{D^{**}}}}
((M_B-M_{D^{**}})u_+(q^2)+(M_B+M_{D^{**}})u_-(q^2))\\
\\
\\
 0           & = & b_+(q^2)  + b_-(q^2)\\
\\
             & = & s_+(q^2) + s_-(q^2) +\sqrt{2}(c_+(q^2) + c_-(q^2))\\
\\
             & = & (M_B+M_{D^{**}})u_+(q^2) + (M_B-M_{D^{**}})u_-(q^2)\\
\\
\end{array}
\end{displaymath}
With the explicit form of the form--factors  given in \cite{isgur2} and
identifying in the heavy mass limit $M_X = m_x = (m_x + m_u)$ where $x$ can be
either $b$ or $c$, eq.(15) follows.

\end{document}